\documentclass{PoS}
\usepackage{amsmath}
\usepackage{subfig}
\usepackage{multicol}
\usepackage{graphicx}

\title{Neutral meson oscillations in the Standard Model and beyond from $\mathbf{N_f=2}$ Twisted Mass Lattice QCD}

\ShortTitle{Neutral meson oscillation in the SM and beyond from $N_f=2$ TM LQCD}

\vspace{-0.23cm}
\author{\speaker{N.~Carrasco}, V.~Gim\'enez\\
        Dep. de F\'isica Te\`orica and IFIC, Universitat de Val\`encia-CSIC\\
\vspace{-0.23cm}
}
\vspace{-0.23cm}
\author{P.~Dimopoulos, R.~Frezzotti, D.~Palao, G.~C.~Rossi \\
        Dip. di Fisica, Universit\`a di Roma "Tor Vergata" and INFN-"Tor Vergata"\\
\vspace{-0.23cm}
}
\vspace{-0.23cm}
\author{G.~Herdoiza\\
        Dto. de F\'isica Te\'orica and Instituto de F\'isica Te\'orica UAM/CSIC \\
\vspace{-0.23cm}
 }
\vspace{-0.23cm}
\author{V.~Lubicz, C.~Tarantino\\
        Dip. di Fisica, Universit\`a Roma Tre and INFN-Roma Tre \\
\vspace{-0.23cm}
}
\vspace{-0.23cm}
\author{G.~Martinelli \\
		SISSA and INFN, Sezione di Roma\\
\vspace{-0.23cm}
}
\vspace{-0.23cm}
\author{M.~Papinutto\\
         Dipartimento di Fisica,Universit\`a di Roma "La Sapienza"  \\
\vspace{-0.23cm}
 }       
\vspace{-0.23cm}
\author{F.~Sanfilippo\\
       Laboratoire de Physique The\'orique, Universit\'e Paris Sud
\vspace{-0.23cm}
}     
\vspace{-0.23cm}                
\author{A.~Shindler\\
        Humboldt-Universitat zu Berlin\\
\vspace{-0.23cm}
}
\vspace{-0.23cm}
\author{S.~Simula\\
        INFN-Roma Tre \\
\vspace{-0.23cm}
}
\vspace{-0.23cm}

\abstract{

We present the ETMC results for the bag parameters describing the neutral kaon mixing in the Standard Model and beyond and preliminary results for the bag parameters controlling the short distance contributions in the $D^0-\overline{D}^0$ oscillations. We also present preliminary results for the $B_{Bd}$, $B_{Bs}$, $B_{Bs}/B_{Bd}$ and $\xi$ parameter controlling  $B^0_{d/s}-\overline{B}^0_{d/s}$ oscillations in the Standard Model employing the so-called ratio method. Using  $N_f=2$ maximally twisted  sea quarks and Osterwalder-Seiler valence quarks  we achieve both $\mathcal{O}(a)$-improvement and continuum like renormalization pattern. Simulations are performed at three-values of the lattice spacing and several values of quark masses in the light, strange, charm region and above charm up to $\sim 2.5 m_c$. Our results are extrapolated to the continuum limit and extrapolated/interpolated to the physical quark masses. 

}

\FullConference{The 30 International Symposium on Lattice Field Theory - Lattice 2012,\\
		June 24-29, 2012\\
		Cairns, Australia}

\begin{document}

\section{Introduction}
Flavour Changing Neutral Currents (FCNC) and CP violation may provide relevant information on the impact of beyond the Standard Model (BSM). The lattice computation of the relevant matrix elements appearing in  $K^0-\overline{K}^0$ and $B^0-\overline{B}^0$ mixing in combination with the experimental value of $\epsilon_K$, $\Delta M_{Bd}$ and $\Delta M_{Bs}$ offers the opportunity to constrain the BSM model parameters. $D^0-\overline{D}^0$, at variance with $K^0-\overline{K}^0$ and $B^0-\overline{B}^0$ mixing, involves valence up-type quarks  so it is sensitive to a different sector of New Physics (NP). 
Although long distance effects dominate over short distance effects in the D system, it is still possible to put significant constrains on the NP parameter space. 
The most general BSM $\Delta F=2$ effective Hamiltonian of dimension-six operators contributing to neutral meson mixing is
\vspace{-0.2cm}
\begin{equation}
H_{\textrm{eff}}^{\Delta F=2}={\displaystyle \sum_{i=1}^{5}C_{i}(\mu)Q_{i}}
\label{eq:Heff}
\end{equation}
where the operators $Q_i$ involving light ($l$) and non-light ($h$) quarks read
\begin{equation}
\begin{array}{lll}
Q_{1}=\left[\bar{h}^{a}\gamma_{\mu}(1-\gamma_{5})l^{a}\right]\left[\bar{h}^{b}\gamma_{\mu}(1-\gamma_{5})l^{b}\right]\\
Q_{2}=\left[\bar{h}^{a}(1-\gamma_{5})l^{a}\right]\left[\bar{h}^{b}(1-\gamma_{5})l^{b}\right] &  & Q_{3}=\left[\bar{h}^{a}(1-\gamma_{5})l^{b}\right]\left[\bar{h}^{b}(1-\gamma_{5})l^{a}\right]\\
Q_{4}=\left[\bar{h}^{a}(1-\gamma_{5})l^{a}\right]\left[\bar{h}^{b}(1+\gamma_{5})l^{b}\right] &  & Q_{5}=\left[\bar{h}^{a}(1-\gamma_{5})l^{b}\right]\left[\bar{h}^{b}(1+\gamma_{5})l^{a}\right]

\label{eq:Qi}
\end{array} 
\end{equation}
\vspace{-0.5cm}
\section{Lattice setup}
Our lattice computations has been performed at three values of the lattice spacing using the $N_f=2$ dynamical quark configurations produced by  ETMC. In the gauge sector, the tree-level Symanzik improved action has been used while the dynamical sea quarks have been regularized employing a twisted mass doublet  at maximal twist \cite{Baron:2009wt}  which provides automatic $\mathcal{O}(a)$ im\-pro\-ve\-ment \cite{Frezzotti:2003ni}. 
The fermionic action for the light doublet $\psi$ in the sea reads in the so-called physical basis
\begin{equation}
S_{\textrm{sea}}^{\textrm{Mtm}}=\sum_{x}\bar{\psi}(x)\left\{ \dfrac{1}{2}\gamma_{\mu}\left(\nabla_{\mu}+\nabla_{\mu}^{*}\right)-i\gamma_{5}\tau^{3}\left[M_{\textrm{cr}}-\dfrac{a}{2}\sum_{\mu}\nabla_{\mu}^{*}\nabla_{\mu}\right]+\mu_{\textrm{sea}}\right\} \psi(x)
\end{equation}
\vspace{-0.15cm}

In addition, both $\mathcal{O}(a)$ improvement and continuum-like renormalization pattern for the four-fermion operators are achieved by introducing an Osterwalder-Seiler \cite{Osterwalder:1977pc} valence quark action allowing for a replica of the heavy $(h,h')$ and the light $(l,l')$ quarks \cite{Frezzotti:2004wz}. The valence quark action reads
\vspace{-0.15cm}
\begin{equation}
S_{\textrm{val}}^{\textrm{OS}}=\sum_{x}\sum_{f=l,l',h,h'}\bar{q}_{f}\left\{ \dfrac{1}{2}\gamma_{\mu}\left(\nabla_{\mu}+\nabla_{\mu}^{*}\right)-i\gamma_{5}r_{f}\left[M_{\textrm{cr}}-\dfrac{a}{2}\sum_{\mu}\nabla_{\mu}^{*}\nabla_{\mu}\right]+\mu_{f}\right\} q_{f}(x) 
\end{equation}
\vspace{-0.1cm}
where the Wilson parameters are conveniently chosen such that $r_h=r_l=r_{h'}=-r_{l'}$.

In table \ref{tab:simulation-detail} we give the details of the simulation and the values of the sea and the valence quark masses at each value of the gauge coupling. The smallest sea quark mass corresponds to a pion of about 280 MeV for the case of $\beta=3.90$. We simulate three heavy valence quark masses $\mu_{``s"}$ around the physical strange one to allow for a smooth interpolation, further three $\mu_{``c"}$ around the physical charm mass followed by a sequence of heavier masses in the range $(m_c, 2.5 m_c)$.
For the inversions in the valence sector we used the stochastic method with propagator sources located at random timeslices in order to increase the statistical information \cite{Foster:1998vw, McNeile:2006bz}. Gaussian smeared quark fields \cite{Gusken:1989qx} are used for masses above the physical charm one \footnote{see the last two columns in table \ref{tab:simulation-detail}} in order to improve the determination of the ground state contribution with respect to the case of simple local interpolating fields.  The value of the smearing parameters are $k_{G}=4$ and $N_{G}=30$. In addition, we apply APE-smearing to the gauge links in the interpolating fields \cite{Albanese:1987ds} with the parameters $\alpha_{APE}=0.5$ and $N_{APE}=20$.
\vspace{-1cm}

\begin{center}
\begin{table}
\centering{} 
\begin{tabular}{cclccc}
\hline 
{\scriptsize $\beta$} & {\scriptsize $a^{-4}(L^{3}\times T)$} & {\scriptsize $a\mu_{l}=a\mu_{\textrm{sea}}$} & {\scriptsize $a\mu_{"s"}$} & {\scriptsize $a\mu_{"c"}$} & {\scriptsize $a\mu_{h}$}\tabularnewline
\hline 
{\scriptsize 3.80} & {\scriptsize $24^{3}\times48$} & {\scriptsize 0.0080 0.0110} & {\scriptsize 0.0165 0.0200 0.0250} & {\scriptsize 0.1982}\textcolor{black}{\scriptsize{} 0.2331 }{\scriptsize 0.2742} & {\scriptsize 0.3225 0.3793 0
.4461 }\tabularnewline
{\scriptsize ($a\sim0.1\mbox{fm})$} &  &  &  &  & {\scriptsize 0.5246 0.6170 }\tabularnewline
\hline 
{\scriptsize 3.90} & {\scriptsize $24^{3}\times48$} & {\scriptsize 0.0040 0.0064} & {\scriptsize 0.0150 0.0220 0.0270} & {\scriptsize 0.1828}\textcolor{black}{\scriptsize{} 0.2150}\textcolor{blue}{\scriptsize{}
}{\scriptsize 0.2529} & {\scriptsize 0.2974 0.3498 0.4114}\tabularnewline
 &  & {\scriptsize 0.0085 0.0100} &  &  & {\scriptsize{} 0.4839 0.5691}\tabularnewline
{\scriptsize $(a\sim0.085\mbox{fm})$} & {\scriptsize $32^{3}\times64$} & {\scriptsize 0.0030 0.0040} &  &  & \tabularnewline
\hline 
{\scriptsize 4.05} & {\scriptsize $32^{3}\times64$} & {\scriptsize 0.0030 0.0060 } & {\scriptsize 0.0120 0.0150 0.0180} & {\scriptsize 0.1572}\textcolor{blue}{\scriptsize{} }\textcolor{black}{\scriptsize 0.1849}\textcolor{blue}
{\scriptsize{}
}{\scriptsize 0.2175} & {\scriptsize 0.2558 0.3008 0.3538 }\tabularnewline
{\scriptsize $(a\sim0.065\mbox{fm})$} &  & {\scriptsize 0.0080} &  &  & {\scriptsize 0.4162 0.4895}\tabularnewline
\hline
\end{tabular}

\caption{\label{tab:simulation-detail}Simulation details} 
\vspace{-0.4cm}
\end{table} 
\end{center}

\section{Bag parameters}

The bag parameters associated to  the operators in Eq.(\ref{eq:Qi}) are defined as
\vspace{-0.15cm}
\begin{equation}
\begin{array}{l}
\langle\overline{P}^{0}|Q_{1}(\mu)|P^{0}\rangle=C_{1}B_{1}(\mu)m_{P}^{2}f_{P}^{2}\equiv C_{1}B_{K}(\mu)m_{P}^{2}f_{P}^{2}\\
\langle\overline{P}^{0}|Q_{i}(\mu)|P^{0}\rangle=C_{i}B_{i}(\mu)\left[\dfrac{m_{P}^{2}f_{P}}{m_{h}(\mu)+m_{l}(\mu)}\right]^{2}
\end{array}
\end{equation}
where $C_i={8/3,-5/3,1/3,2,2/3}$, $i=1,..,5$.
The renormalisation constants (RCs) of the relevant four- and two-fermion operators has been computed non-per\-turba\-tively in the RI'-MOM scheme \cite{Martinelli:1994ty}. Due to the OS-tm mixed action setup, the renormalized values of the bag parameters are given by the formulae \cite{Frezzotti:2004wz} \cite{Bertone:2012cu} \cite{Constantinou:2010qv}
\vspace{-0.15cm}
\begin{equation}
\begin{array}{ccc}
\hat{B}_{1}=\dfrac{Z_{11}}{Z_{A}Z_{V}}B_{1}, & \,\,\,\,\,\,\,\,\,\,\,\,\,\,\,\,\,\,\,\,\,\, & \hat{B}_{i}=\dfrac{Z_{ij}}{Z_{P}Z_{s}}B_{j}\,\,\,\,\,\,\,\,\,\,\,\,\, i,j=2,..,5\end{array}
\end{equation}
\vspace{-0.8cm}
\section{$K^0-\overline{K}^ 0$}
For large time separation $y_0\ll x_0 \ll y_0+T_{\textrm{sep}}$ the plateau of the following ratio estimates
\vspace{-0.15cm}
\begin{equation}
\begin{array}{ccc}
E[B_{1}](x_{0})=\dfrac{C_{1}(x_{0})}{C_{AP}(x_{0})C'_{AP}(x_{0})}, &  \,\,\,\,\,\,\,\,\,\,\,\,\,\,\,\,\,\,\,\,\,\, & E[B_{i}](x_{0})=\dfrac{C_{i}(x_{0})}{C_{PP}(x_{0})C'_{PP}(x_{0})}\end{array}
\end{equation}
 \vspace{-0.1cm}
provides the estimate of bare $B_i$. The involved correlators are
\vspace{-0.1cm}
\begin{equation}
\begin{array}{lcl}
C_{i}(x_{0})={\displaystyle \sum_{\vec{x}}}\langle\mathcal{P}_{y_{0}+T_{\textrm{sep}}}^{43}Q_{i}(\vec{x},x_{0})\mathcal{P}_{y_{0}}^{21}\rangle\\
C_{PP}(x_{0})={\displaystyle \sum_{\vec{x}}}\langle P^{12}(\vec{x},x_{0})\mathcal{P}_{y_{0}}^{21}\rangle , &  & C_{AP}(x_{0})={\displaystyle \sum_{\vec{x}}}\langle A^{12}(\vec{x},x_{0})\mathcal{P}_{y_{0}}^{21}\rangle\\
C'_{PP}(x_{0})={\displaystyle \sum_{\vec{x}}}\langle\mathcal{P}_{y_{0}+T_{\textrm{sep}}}^{43}P^{34}(\vec{x},x_{0})\rangle , &  & C'_{AP}(x_{0})={\displaystyle \sum_{\vec{x}}}\langle\mathcal{P}_{y_{0}+T_{\textrm{sep}}}^{43}A^{34}(\vec{x},x_{0})\rangle
\end{array}
\end{equation}
\vspace{-0.1cm}
with \footnote{$\mathcal{P}$ sources are local for K-meson and smeared for D- and B- mesons}   $\begin{array}{ccc}
\mathcal{P}_{y_{0}}^{21}={\displaystyle \sum_{\vec{y}}\bar{q}_{2}(\vec{y},y_{0})\gamma_{5}q_{1}(\vec{y},y_{0})}, &  & \mathcal{P}_{y_{0}}^{43}={\displaystyle \sum_{\vec{y}}\bar{q}_{4}(\vec{y},y_{0}+T_{\mbox{\textrm{sep}}})\gamma_{5}q_{3}(\vec{y},y_{0}+T_{\textrm{sep}})}\end{array}$ and $P^{ij}=\bar{q}_i\gamma_5q_{j}$, $A^{ij}=\bar{q}_i\gamma_0\gamma_5q_{j}$. In table \ref{tab:BK-results} we gather our final continuum results for $B_i$ in the $\overline{MS}$ scheme. For details about the analysis, results and its phenomenological implications we refer to \cite{Bertone:2012cu} and \cite{Constantinou:2010qv}.

\vspace{-0.6cm}
\begin{center}
\begin{table}
\centering{} 
\begin{footnotesize}
\begin{tabular}{|c|c|c|c|c|}
\hline 
\multicolumn{5}{|c|}{$\overline{MS}$(2 GeV)}\tabularnewline
\hline 
\hline 
$B_{1}$ & $B_{2}$ & $B_{3}$ & $B_{4}$ & $B_{5}$\tabularnewline
\hline 
0.52(02) & 0.54(03) & 0.94(08) & 0.82(05) & 0.63(07)\tabularnewline
\hline 
\end{tabular}
\end{footnotesize}
\vspace{-0.2cm}
\caption{\label{tab:BK-results}Continuum limit results for $B_i$ parameters of the $K^0-\overline{K}^0$ system renormalized in the $\overline{MS}$  scheme  of \cite{Buras:2000if} at 2GeV using M1-type RCs defined in \cite{Constantinou:2010qv}} 
\vspace{-0.4cm}
\end{table} 
\end{center}

\vspace{-1cm}
\section{$D^0-\overline{D}^ 0$}
\vspace{-0.2cm}
The bag parameters for the $D^ 0-\overline{D}^0$ oscillations can be determined following the same strategy as in \cite{Bertone:2012cu} and outlined in the previous section, now with two of the OS valence quarks representing the up quark while the other two will be identified with the charm quark, i.e $l \sim u$ and $h \sim c$.  Physical values are obtained by interpolating data in $\mu_{``c"}$  to the physical value $\mu_{c}$  while chiral and continuum extrapolations are carried out simultaneously. The physical values for the charm quark mass have been previously derived and can be found in \cite{Blossier:2010cr}.

\begin{figure}
  \centering
  \subfloat[]{\label{fig:B1-plateau-sme}\includegraphics[scale=0.17]{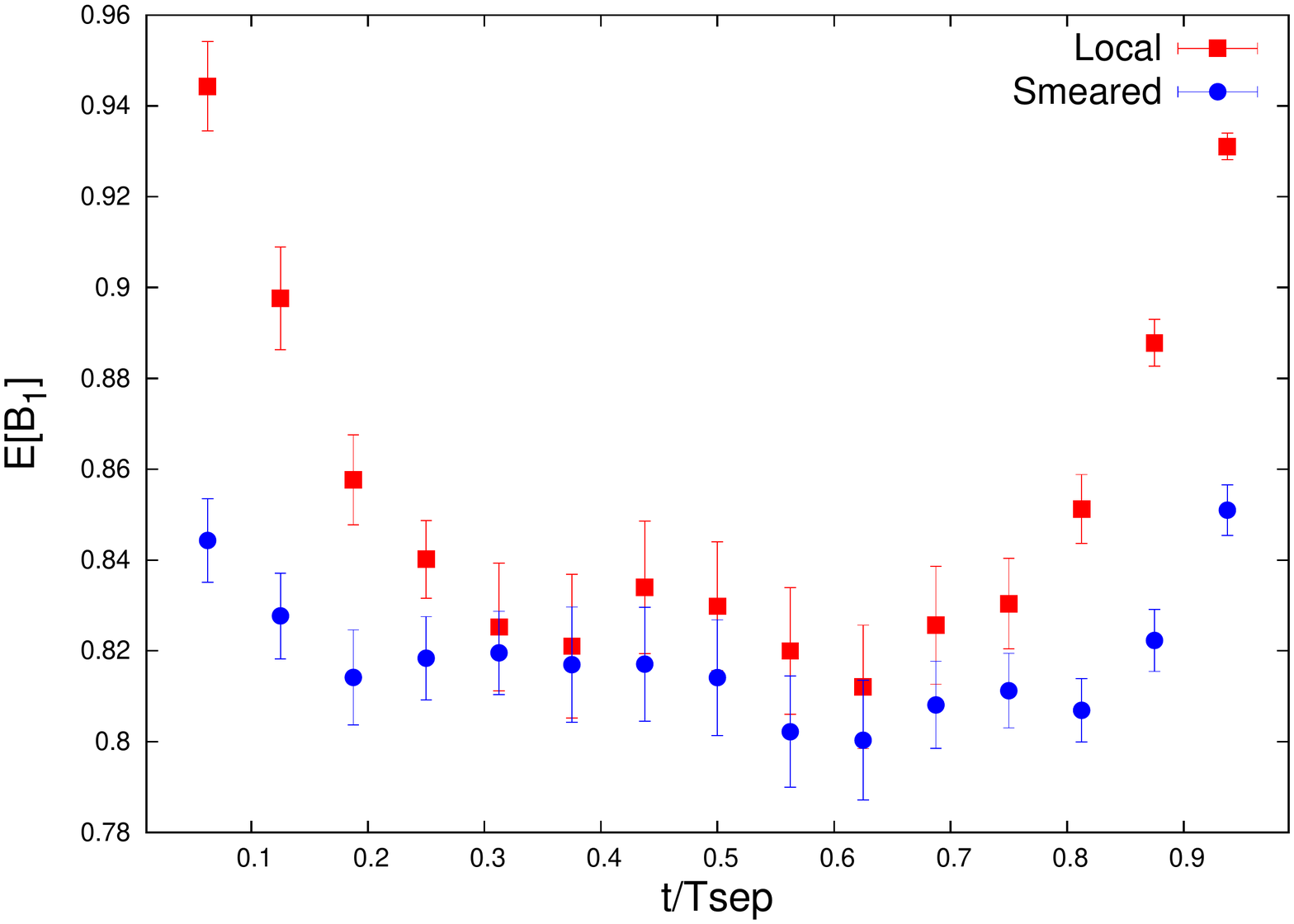}}\hspace{2cm}
  \subfloat[]{\label{fig:B-plateau}\includegraphics[scale=0.17]{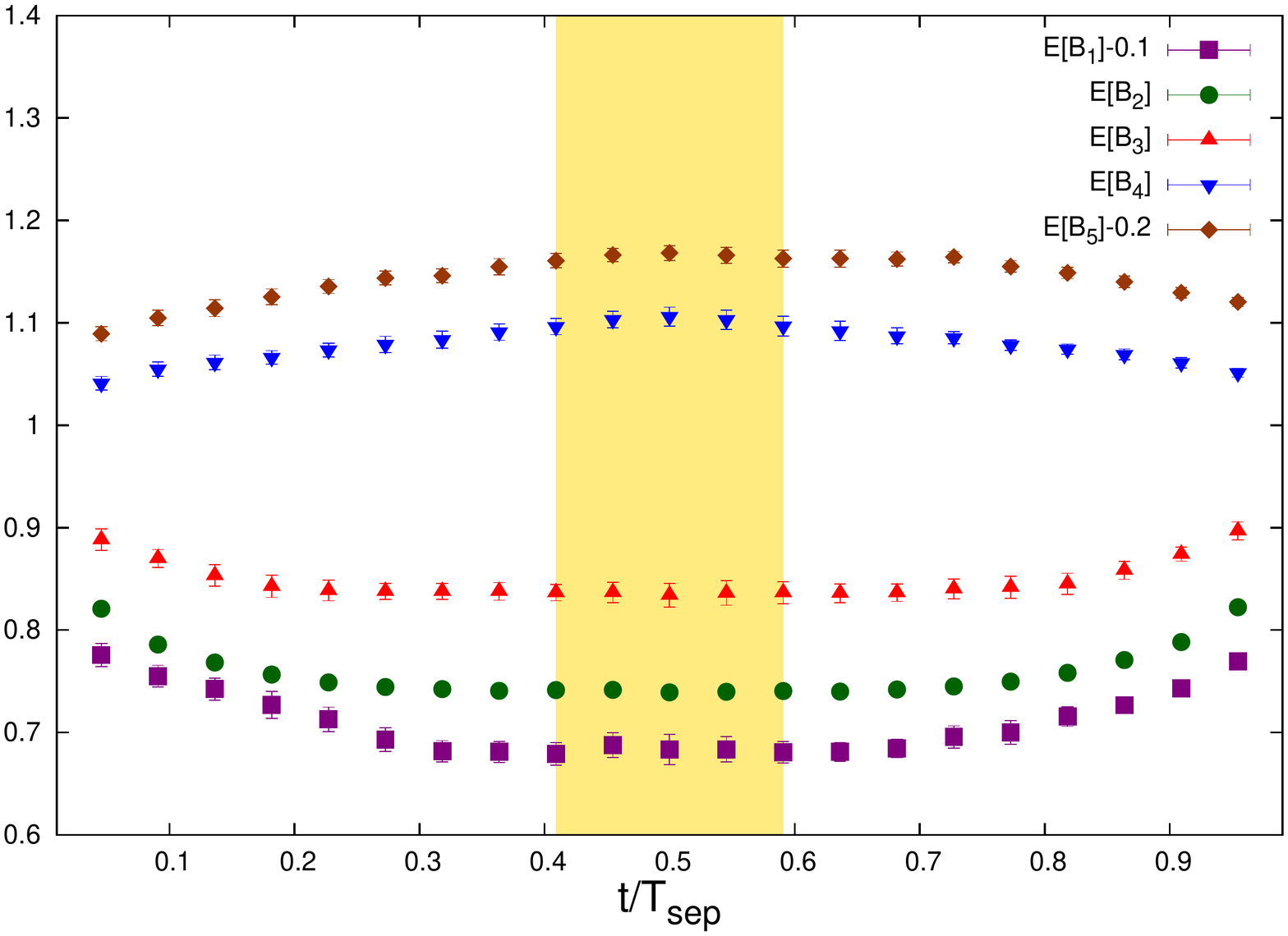}}
\vspace{-0.2cm}
  \caption{(a) Comparison of the plateau for the estimator of the $B_1$ bag parameter at $\beta=3.80$  on a $24^{3}\times48$  lattice and $(a\mu_{l},a\mu_{h})=(0.0080,0.2331)$ with local and smeared sources (b) Plateaux for $E[B_i]$ (i=1,...,5)  plotted vs $t/T_{\textrm{sep }}$ for  $\beta=4.05$ , $(a\mu_{l},a\mu_{h})=(0.0030,0.1849)$  on a $32^{3}\times64$  lattice. The shaded region delimits the plateau interval. }
\vspace{-0.2cm}
\end{figure}

Exploratory studies show that using Gaussian smeared sources and choosing a time separation between meson sources smaller than $T/2$ (we set $T_{\textrm{sep}}=16$ at $\beta=3.8$, $T_{\textrm{sep}}=18$ at $\beta=3.9$, $T_{\textrm{sep}}=22$ at $\beta=4.05$) is crucial for quark masses around the physical charm and above. This improvement allows us to extract the ground state with more confidence and precision in a wider time interval.  From figure \ref{fig:B1-plateau-sme}  the benefit of Gaussian smearing compared to  local source and sink is evident. For illustration in figure \ref{fig:B-plateau} we display the quality plateau of $B_i$ at the smallest lattice spacing, $\beta=4.05$, and for the smallest value of the light quark. Preliminary results are collected in table \ref{tab:BD-results}.

\vspace{-0.15cm}

\begin{center}
\begin{table}
\centering{} 

\begin{footnotesize}
\begin{tabular}{|c|c|c|c|c|}

\hline 
\multicolumn{5}{|c|}{$\overline{MS}$(2 GeV)}\tabularnewline
\hline 
\hline 
$B_{1}$ & $B_{2}$ & $B_{3}$ & $B_{4}$ & $B_{5}$\tabularnewline
\hline 
0.78(04) & 0.71(04) & 1.10(12) & 0.94(06) & 1.17(15)\tabularnewline
\hline 

\end{tabular}
\end{footnotesize}
\vspace{-0.25cm}

\caption{\label{tab:BD-results}Continuum limit results for $B_i$ parameters of the $D^0-\overline{D}^0$ system renormalized in the $\overline{MS}$ scheme of \cite{Buras:2000if}  at 2GeV using M1-type RCs defined in \cite{Constantinou:2010qv}} 
\vspace{-0.2cm}
\end{table} 
\end{center}
\vspace{-1cm}
\section{$B^0-\overline{B}^0$}

Since discretization errors on current lattices are expected to be large at the physical value of b-quark mass, our strategy for the computation of the bag-parameters in the B sector is  based on the ratio method approach proposed in \cite{Blossier:2009hg} by introducing suitable ratios with exactly known static limit and interpolating them to the b-quark mass.   

A similar strategy to the one employed in \cite{Dimopoulos:2011gx} for the computation of the $B_d$ an $B_s$ decay constants is followed here for the $B_{Bd}$ and $B_{Bs}$ parameters. 
According to HQET, the ratio of renormalized bag parameters evaluated in QCD is expected to approach unity as $1/\hat{\mu}_h \rightarrow 0$. The leading deviation are predicted to be of order $1/\log(\hat{\mu}_h/\Lambda_{QCD})$ and are expected to be tiny in the $\hat{\mu}_h$ range of our data. We consider the following  chiral and continuum extrapolated scalar ratios, where the corrections to the power scaling in $1/\hat{\mu}_h$ (which are found
to be at most $\sim$ one standard deviation, see Eq.(\ref{eq:BBds})) are estimated in perturbation theory by matching HQET to QCD through the C-factors

\vspace{-0.15cm}

\begin{equation}
\begin{array}{l}
\omega_{d}\left(\hat{\mu}_{h}^{(n)}\right)=\lim_{\hat{\mu}_{sea}\rightarrow\hat{\mu}_{u/d}}\lim_{a\rightarrow0}\omega_{d}^{L}\left(\hat{\mu}^{(n)};\hat{\mu}_{sea},\hat{\mu}_{l},a\right),\\
\omega_{s}\left(\hat{\mu}_{h}^{(n)}\right)=\lim_{\hat{\mu}_{sea}\rightarrow\hat{\mu}_{u/d}}\lim_{a\rightarrow0}\omega_{s}^{L}\left(\hat{\mu}^{(n)};\hat{\mu}_{sea},\hat{\mu}_{s},a\right),\end{array} 
\end{equation}
\vspace{-0.35cm}
with
\vspace{-0.35cm}
\begin{equation}
\begin{array}{lcl}
\omega^{L}_{d/s}\left(\mu_{h}^{(n)};\mu_{sea},\hat{\mu}_{l},a\right) & = & \dfrac{C\left(\hat{\mu}_{h}^{(n)};\hat{\mu}^{*},\mbox{\ensuremath{\mu}}\right)}{C\left(\hat{\mu}_{h}^{(n)}/\lambda;\hat{\mu}^{*},\mu\right)}\dfrac{B_{Bd/s}\left(\hat{\mu}_{h}^{(n)};\hat{\mu}_{sea},\hat{\mu}_{l/s},a\right)}{B_{Bd/s}\left(\hat{\mu}_{h}^{(n)}/\lambda;\hat{\mu}_{sea},\hat{\mu}_{l/s},a\right)}\end{array} 
\label{eq:scaling-law}
\end{equation}

\vspace{-0.25cm}

The C-factors ratio contains the information on the $1/\log(\hat{\mu}_h)$ corrections. At tree-level (TL)  $C\left(\hat{\mu}_{h}^{(n)};\hat{\mu}^{*},\mbox{\ensuremath{\mu}}\right)=1$, while at leading-log (LL) it is given by $$\ensuremath{C\left(\hat{\mu}_{h}^{(n)};\hat{\mu}^{*},\mbox{\ensuremath{\mu}}\right)}=\left[\alpha(\hat{\mu}^{*})/\alpha(\mu_{h}^{(n)})\right]^{-(\tilde{\gamma}_{0}^{11})/2\beta_{0}}\left[\alpha(\mu_{h}^{(n)})/\alpha(\mu)\right]^{-(\gamma_{0}^{11})/2\beta_{0}}\left[\alpha(\hat{\mu}^{*})/\alpha(\mu_{h}^{(n)})\right]^{\tilde{\gamma}_{A}/\beta_{0}}$$
and at NLL (not included in this preliminary analysis) the HQET mixing of $O_1$ with $O_{2,3}$ should be considered. In the formula above $\tilde{\gamma}_A$ and $\tilde{\gamma}_0^{11}$ characterize the anomalous dimension (AD) of the axial $hl$-current and the ``11'' element of the AD-matrix of the $hlhl$ operator in HQET.

In order to have better control on the chiral extrapolation, we consider the double ratio $B_{Bs}/B_{Bd}$
\vspace{-0.15cm}
\begin{equation}
\zeta_{\omega}^{L}\left(\hat{\mu}^{(n)};\hat{\mu}_{sea},\hat{\mu}_{l},a\right)=\dfrac{\omega^{L}_s\left(\hat{\mu}^{(n)};\hat{\mu}_{sea},\hat{\mu}_{s},a\right)}{\omega^{L}_d\left(\hat{\mu}^{(n)};\hat{\mu}_{sea},\hat{\mu}_{l},a\right)}
\label{zeta-omega}
\end{equation}
which also tends to 1 in the continuum and chiral limit. The ratio method is also applied to the phenomenologically interesting quantity $\xi=(f_{Bs}/f_{Bd})\sqrt{B_{Bs}/B_{Bd}}$ by forming the ratios $\zeta_{\xi}^L$ analogous to Eq.(\ref{zeta-omega}) at successive values of the heavy quark mass.

\begin{figure}
\vspace{-0.5cm}
\begin{center}
   \subfloat[]{\label{fig:chiral-cont-ex}\includegraphics[scale=0.2]{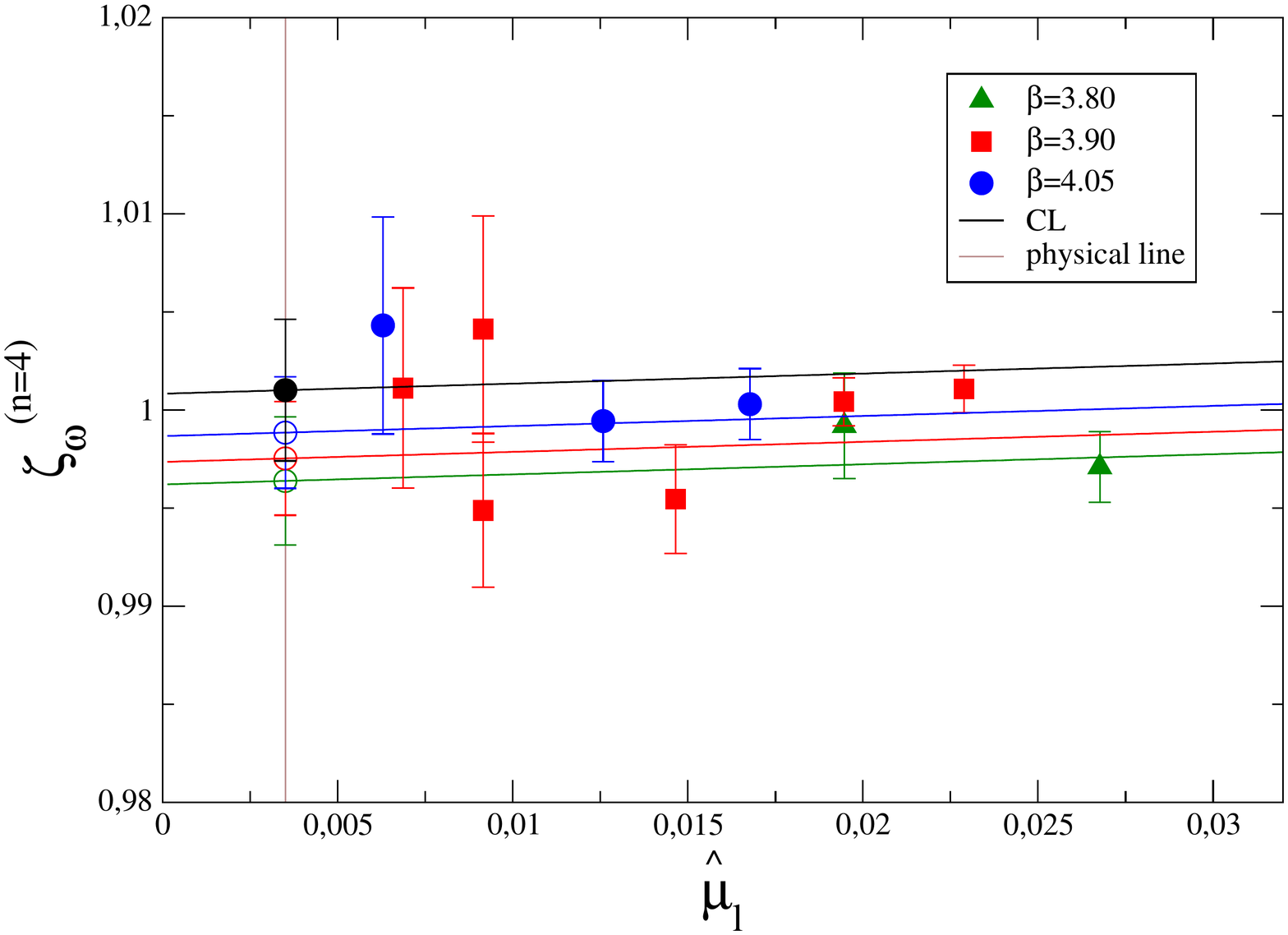}}    
      \subfloat[]{\label{fig:chiral-cont-ex-trigg}\includegraphics[scale=0.2]{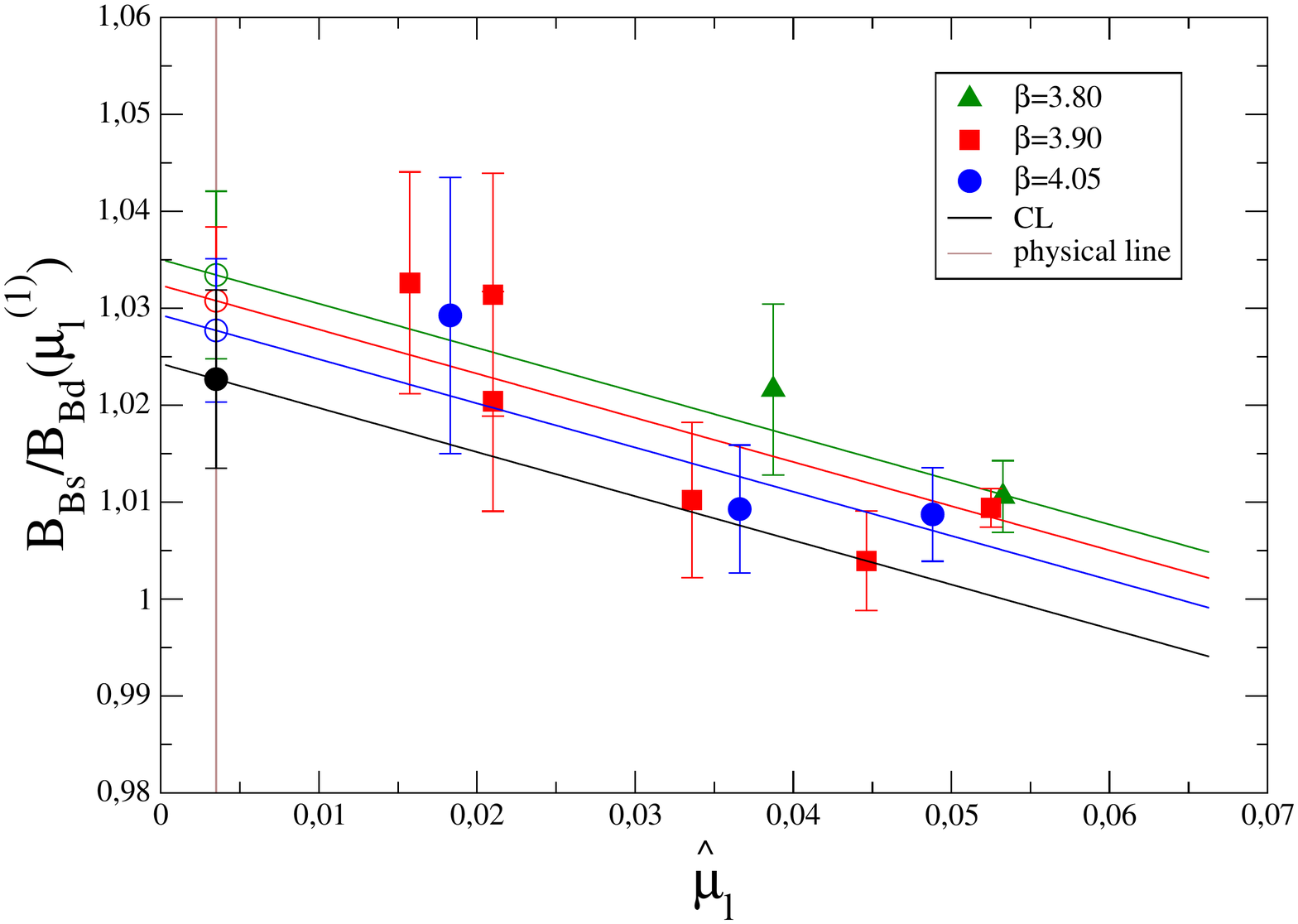}}   \hspace{2cm}
   
  \vspace{-0.4cm}
 \caption{Double ratio $\zeta_{\omega}$ for the fourth  analysed mass (a) Chiral and continuum extrapolation of the ratio $B_{Bs}/B_{Bd}$ computed at the triggering point  (b). Vertical lines represent the position of the physical point }
 \end{center}
\end{figure}

The quantities $\omega_d^L$, $\omega_s^L$, $\zeta_{\omega}^L$  and $\zeta_{\xi}^L$ have a smooth chiral and continuum extrapolation, showing no significant dependence on $\mu_l$ and small cutoff effects. The results turn out to be well described by a linear dependence in $\mu_l$ and $a^2$. For instance in figure \ref{fig:chiral-cont-ex} we show the chiral and continuum extrapolation of $\zeta_{\omega}^L$ at the fourth of the considered masses. 

\begin{figure}  

\vspace{-0.7cm}
  \subfloat[]{\label{fig:zeta_Bs}\includegraphics[scale=0.4, trim=110 465 80 100,clip]{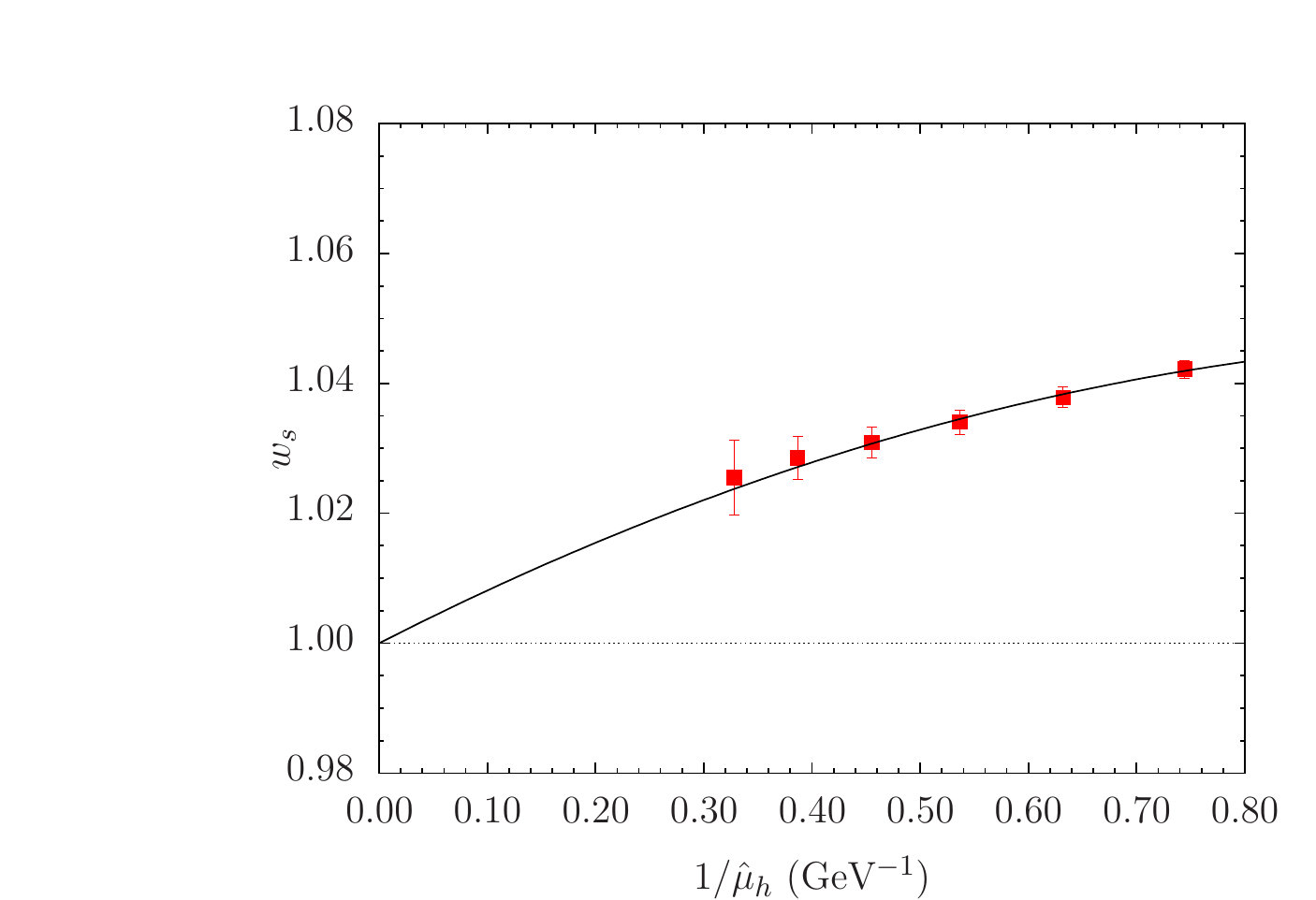}}\hspace{-1.05cm}
  \subfloat[]{\label{fig:zeta_w}\includegraphics[scale=0.4, trim=110 465 80 100,clip]{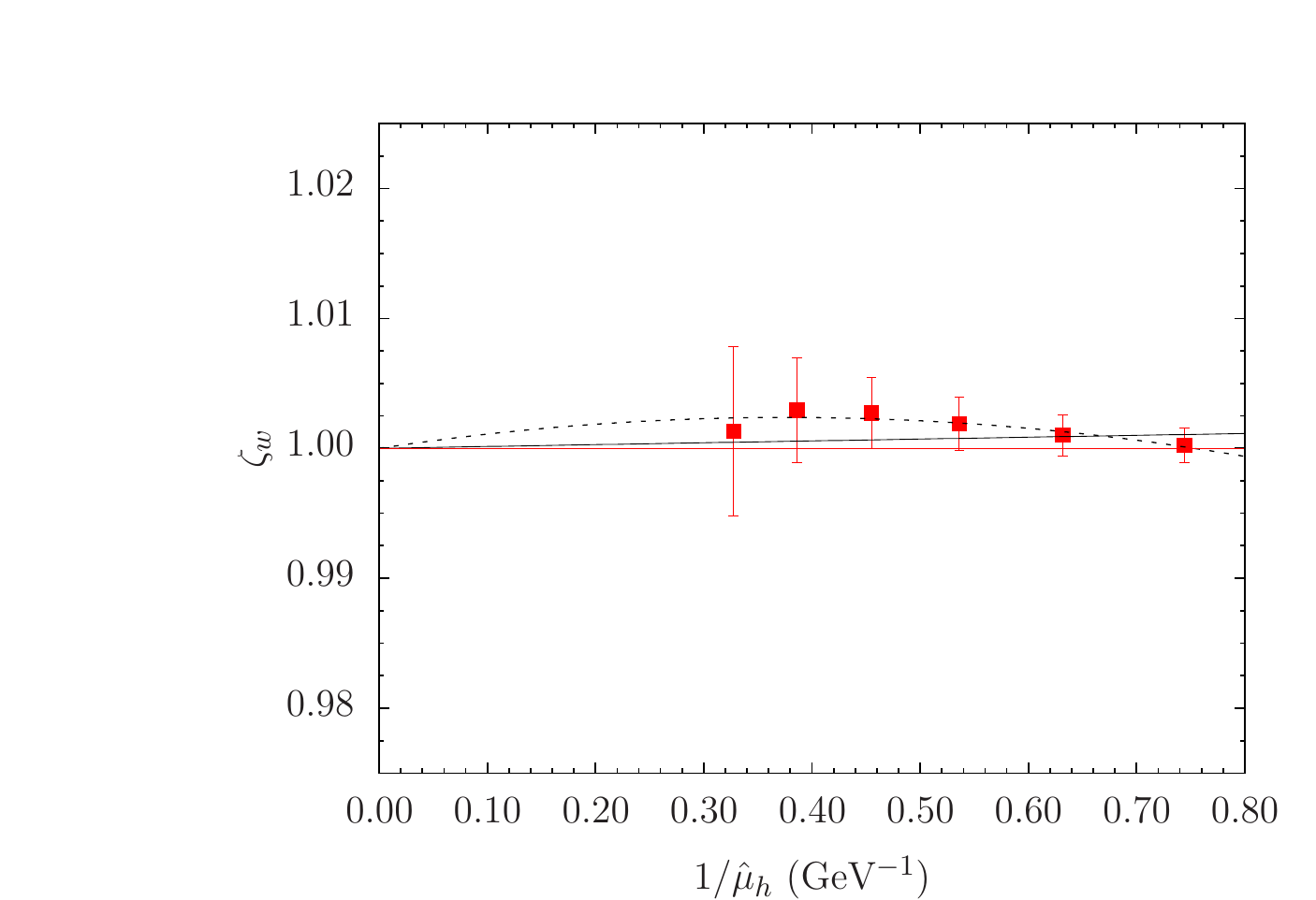}}\hspace{-1.05cm}
  \subfloat[]{\label{fig:zeta_xi}\includegraphics[scale=0.4, trim=110 465 80 100,clip]{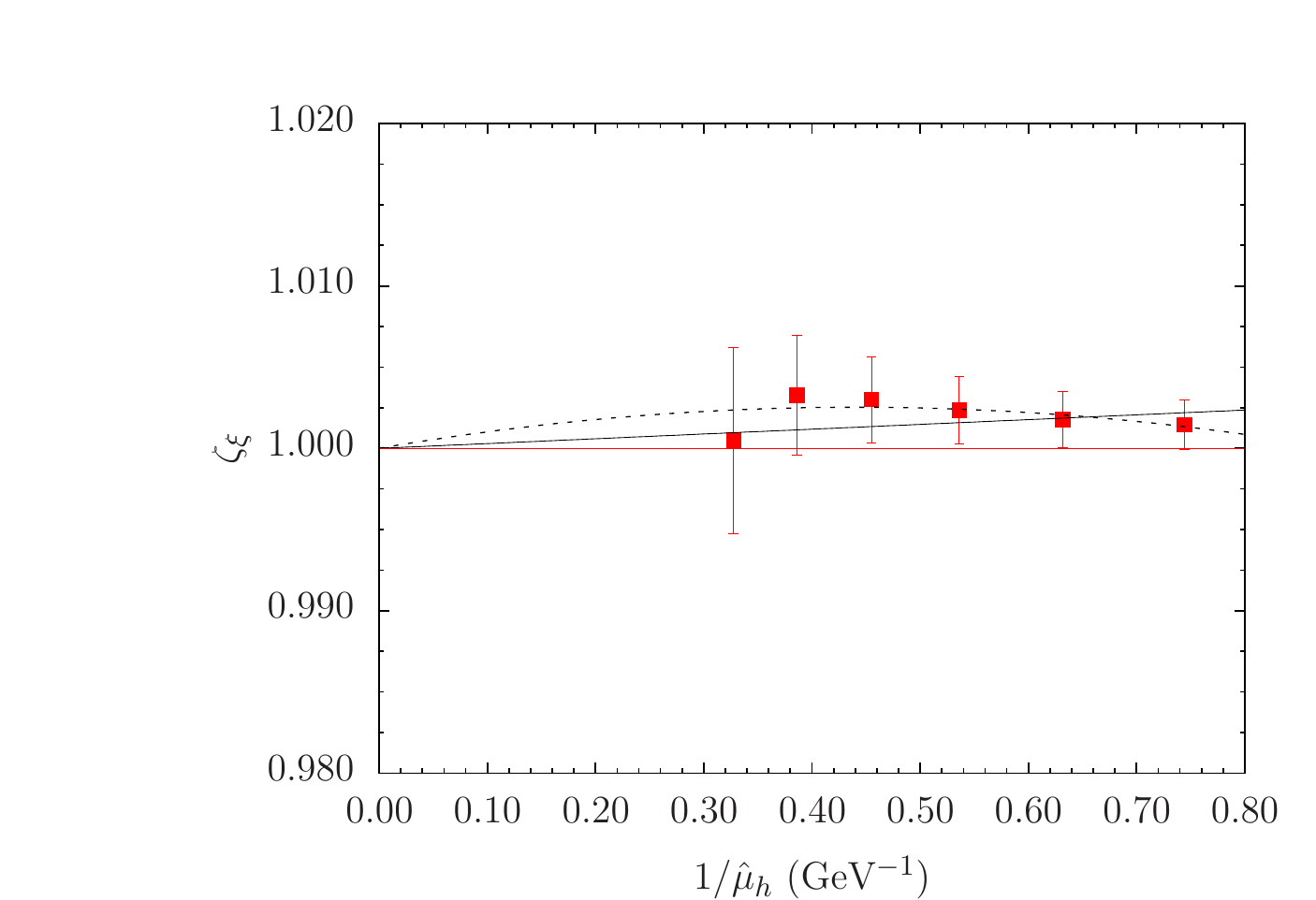}}
  \vspace{-0.35cm}
 \caption{Heavy quark mass dependence of the ratio $\omega_s$ (a) and the double ratios $\zeta_{\omega}$ (b)  and $\zeta_{\xi}$ (c) extrapolated to the physical values of the light and strange masses and to the continuum limit. }

\end{figure}

Finally, we study the dependence of the ratios $\omega_d$, $\omega_s$, $\zeta_{\omega}$  and $\zeta_{\xi}$ on the inverse of the heavy quark mass as shown in figures \ref{fig:zeta_Bs}, \ref{fig:zeta_w} and \ref{fig:zeta_xi} for $\omega_s$, $\zeta_{\omega}$  and $\zeta_{\xi}$. For $\omega_d$ and $\omega_s$ we perform a linear ($\omega_{d/s}=1+b(\lambda)/\hat{\mu}_h$) or a quadratic ($\omega_{d/s}=1+b(\lambda)/\hat{\mu}_h+c(\lambda)/\hat{\mu}_h^2$) interpolation to the b-quark mass and the difference between them will be eventually assigned as a systematic error. In contrast, $\zeta_{\omega}$  and $\zeta_{\xi}$  show very weak dependence on the heavy quark mass, thus in these cases we perform either a quadratic or linear interpolation or we fix the ratio equal to its asymptotic heavy-quark mass limit $\zeta_{\omega}=1$  and $\zeta_{\xi}=1$.

From the structure of Eq.(\ref{eq:scaling-law}) we derive the iterative formulae
\vspace{-0.15cm}
\begin{equation}
\begin{array}{lcl}
\omega_{d,s}\left(\hat{\mu}_{h}^{(2)}\right)\omega_{d,s}\left(\hat{\mu}_{h}^{(3)}\right)...\omega_{d,s}\left(\hat{\mu}_{h}^{(K+1)}\right) & = & \dfrac{C\left(\hat{\mu}_{h}^{(K+1)}\right)B_{Bd,s}\left(\hat{\mu}_{h}^{(K+1)}\right)}{C\left(\hat{\mu}_{h}^{(1)}\right)B_{Bd,s}\left(\hat{\mu}_{h}^{(1)}\right)},\\
\zeta_{\omega}\left(\hat{\mu}_{h}^{(2)}\right)\zeta_{\omega}\left(\hat{\mu}_{h}^{(3)}\right)...\zeta_{\omega}\left(\hat{\mu}_{h}^{(K+1)}\right) & = & \dfrac{B_{Bs}/B_{Bd}\left(\hat{\mu}_{h}^{(K+1)}\right)}{B_{Bs}/B_{Bd}\left(\hat{\mu}_{h}^{(1)}\right)},\\
\zeta_{\xi}\left(\hat{\mu}_{h}^{(2)}\right)\zeta_{\xi}\left(\hat{\mu}_{h}^{(3)}\right)...\zeta_{\xi}\left(\hat{\mu}_{h}^{(K+1)}\right) & = & \dfrac{f_{Bs}/f_{Bd}\sqrt{B_{Bs}/B_{Bd}}\left(\hat{\mu}_{h}^{(K+1)}\right)}{f_{Bs}/f_{Bd}\sqrt{B_{Bs}/B_{Bd}}\left(\hat{\mu}_{h}^{(1)}\right)}
\label{eq:master}
\end{array} 
\end{equation}
relating the quantity at the heavy quark mass $\hat{\mu}_h^{(K+1)}$ with its triggering value at $\hat{\mu}_h^{(1)}$ and the fit values of $\omega_d$, $\omega_s$, $\zeta_{\omega}$  and $\zeta_{\xi}$, where the value of $\hat{\mu}_h^{(1)}$ together with the $\lambda$ factor and the number of steps to arrive to the b quark mass point have been previously tuned  in such a way that after a finite number of steps (K) the heavy-light meson mass assumes the experimental value of $M_B$ \cite{AndreaPOS}.  The value of the quantities in Eq.(\ref{eq:master}) at the triggering mass point can be accurately measured in the chiral and continuum limit since $\mu_h^{(1)}$ lies in the well accessible charm quark mass region. As example, in figure  \ref{fig:chiral-cont-ex-trigg}  we display the continuum and chiral extrapolation of the ratio $B_{Bs}/B_{Bd}$ at triggering mass, which shows small cutoff effects. 
Finally, Eq.(\ref{eq:master}) leads to the following preliminary estimates:

\vspace{-0.35cm}
\begin{equation}
\begin{array}{ccc}
B_{Bd}=0.87(05), & \,\,\,\,\,\,\,\,\,\,\,\,\,\,\,\,\,\,\,\,\,\,\, & B_{Bs}=0.90(05)
\label{eq:BBds}
\end{array}
\end{equation}
where the error is the sum in quadrature of statistical and systematic uncertainties coming from the variation of the result if we use a linear  or quadratic fit in the heavy quark mass and the truncation of the C-factors estimated as the difference between TL and LL matching. C-factors cancel out in the ratios $B_{Bs}/B_{Bd}$ and $\xi$, while the implementation of a constant, a linear or a quadratic fit in $\hat{\mu}_h$ turns out to provide essentially  identical results. The preliminary estimates are

\vspace{-0.35cm}
\begin{equation}
\begin{array}{ccc}
B_{Bs}/B_{Bd}=1.03(2), & \,\,\,\,\,\,\,\,\,\,\,\,\,\,\,\,\,\,\,\,\,\,\,  & \xi=1.21(06)
\end{array} 
\end{equation}
where in the second one we added in quadrature a systematic error due to the fit ansatz at the triggering mass point. 

\vspace{-0.6cm}
\section{Acknowledgements}
\vspace{-0.1cm}
CPU time was provided by the Italian SuperComputing Resource Allocation (ISCRA) under the class A project HP10A7IBG7 "A New Approach to B-Physics on Current Lattices" and the class C project HP10CJTSNF "Lattice QCD Study of B-Physics" at the CINECA supercomputing service. We also acknowledge computer time made available to us by HLRN in Berlin.

\end{document}